# Frequency-dependent complex conductivity of an organic thin-film transistor

Daniel R. Lenski, Adrian Southard, Michael S. Fuhrer

*Department of Physics and Center for Nanophysics and Advanced Materials, University of Maryland, College Park, Maryland 20742-4111, USA*

We measure the complex impedance between source/drain electrodes and the gate electrode of a pentacene thin-film transistor (TFT) at frequencies 50 Hz < $\omega/2\pi$ < 20 kHz. Modeling the TFT as a distributed RC network (RC transmission line), we find that the data cannot be explained by a model including only a real, frequency-independent sheet conductivity. Instead, we use the RC transmission line model to extract the frequency-dependent complex sheet conductivity $\sigma(\omega) = \sigma'(\omega) + j\sigma''(\omega)$ of the pentacene film. At high frequencies, $\sigma(\omega)$ increases with frequency, $\sigma'(\omega)$ and $\sigma''(\omega)$ become similar in magnitude, and the on/off ratio is significantly reduced.

Organic thin-film transistors (OTFTs) are among the most prominent and promising applications of organic semiconductors. However, most studies of these devices' performance have used dc measurements exclusively, while many applications, such as flexible display back-planes[1,2] and radio-frequency ID tags[3,4] require ac operation. Bulk organic semiconductors show strongly frequency-dependent conductivity,[5-8] typical of disordered solids,[9,10] at frequencies in this range, however, we know of only one attempt to extract the intrinsic conductivity of an organic thin film at finite frequency,[11] but this study explicitly assumed no frequency dependence of the film conductivity.[9,10]

In this paper, we report on the dc and ac (50 Hz – 20 kHz) characteristics of polycrystalline pentacene TFTs, adopting a distributed RC network model, specifically an RC transmission-line model,[12] for the ac analysis. We find significant disagreement between the data and a model assuming a frequency-independent real conductivity of the pentacene film. The agreement cannot be improved by including a complex contact resistance or a gate-voltage-dependent interface trap capacitance.[11] Instead, we find that the data can only be explained by a complex frequency-dependent conductivity $\sigma(\omega) = \sigma'(\omega) + j\sigma''(\omega)$, and we extract this conductivity for the pentacene thin film. $\sigma(\omega)$ shows strong frequency dependence, which should have profound implications for design of organic TFTs operating at finite frequency.

We fabricate bottom-contacted pentacene devices on 300 nm thermally grown $SiO_2$ over degenerately doped silicon which serves as the gate electrode. Gold electrodes of 50 nm thickness are deposited by thermal evaporation through a shadow mask. After heating powdered pentacene (Sigma-Aldrich #684848, 99.9%) to just below its sublimation point to remove impurities,[13] the pentacene is thermally evaporated through a shadow mask, depositing at a rate around 0.006 nm/s to a thickness of 23 nm.

After fabrication, the devices are placed in a vacuum probe station ($P < 10^{-6}$ Torr) to maintain stability of the pentacene, where they remain for the duration of our measurements. The dc drain current-drain voltage ($I_D$-$V_{DS}$) characteristic is measured over -5 V < $V_{DS}$ < +5 V, and a linear fit performed to find the dc sheet conductivity is then $\sigma(0) = (L/W)(dI_D/dV_{DS})$. For ac characterization, the measurement circuit is electronically reconfigured to the "transmission line configuration" (see Fig. (2) inset): a set of relays shorts the source and drain contacts together, and these are connected to the high terminal of an Andeen-Hagerling AH2700A capacitance bridge, while the gate is connected to the low terminal. The complex admittance between gate and source/drain contacts $Y(\omega) = G(\omega) + jB(\omega)$ is measured as a function of frequency from 50 Hz < $\omega/2\pi$ < 20 kHz. The ac bias voltage is kept less than 1 V. Note that the conductance measured in the transmission-line configuration is not the conductance from source to drain (notably the transmission-line conductance is zero at $\omega = 0$); at finite frequency the transmission-line conductance results from the propagation of the ac signal into the lossy transmission line consisting of the pentacene film over the gate electrode, as discussed below. In order to minimize discrepancies due to bias stress, we set the gate voltage ($V_G$), then perform the dc measurement followed by the ac measurement, without adjusting the gate voltage in between. Furthermore, we start at most positive $V_G$ and move towards most negative $V_G$.

Figure 1 shows the dc characteristics of a representative device, of channel length 1262 μm and width 1600 μm. The device shows typical *p*-type behavior with a threshold voltage $V_T$ around -28 V, a field-effect mobility of 0.15 cm$^2$/Vs, and an on/off ratio approaching 10$^4$ (the off conductance of ~10 pS is likely not a true measure of the channel conductance but rather a floor set by leakage of the gate oxide, so the on/off ratio may be an underestimate).

Using the model developed by Chow and Wang,[12] and imposing appropriate boundary conditions, we obtain the admittance of the transmission line:

$$Y = G + jB$$
$$G = 2\sigma z \, W/L \, \frac{\sinh 2z - \sin 2z}{\sinh^2 z + \cos^2 z}$$
$$B = 2\sigma z \, W/L \, \frac{\sinh 2z \, \sin 2z}{\sinh^2 z + \cos^2 z} \qquad (1)$$
$$\text{where } z = \sqrt{\omega c / 2\sigma} \, L/2$$

For our first approach, we assume purely real sheet conductivity of the semiconductor, $\sigma$, and capacitance per unit area, $c$ (reasonable for $SiO_2$ at $\omega/2\pi < 20$ kHz). This model predicts admittance depending only on a frequency-independent $c$ and $\sigma$, and with frequency-dependence only through $z$, a unit-less measure of effective device length. At high frequencies $\omega \gg 8/crL^2$ the effective device length is large, $z \gg 1$, and equation (1) simplifies to $G(\omega) = B(\omega) = (\omega c/2r)^{1/2} W$. This is the classic result for an infinite $RC$ transmission line.

Figure 2 shows the ac admittance, $G(\omega)$ and $B(\omega)$, in the transmission line configuration, as well as lines representing the prediction of Eqn. (1) using $\sigma = \sigma(0)$ calculated from the dc conductivity, and $c$ calculated from the oxide thickness and dielectric constant (there are no adjustable parameters in the model). The measured capacitance of the gold source and drain electrodes is subtracted from the measured susceptance $B(\omega)$. The electrodes' capacitance is measured to be 89.59 pF (in excellent agreement with the value calculated from their combined area of 0.8 mm$^2$), with test electrodes showing no variation of capacitance from dc to 20 kHz, no variation with dc bias up to ±75 V, and a loss tangent of < 0.002. There is good agreement between the experimental data and the model at low frequencies for the most negative $V_G$ values, i.e. $V_G = $ -30, -40, -50 V (when the semiconductor channel is on). However, at higher frequencies and higher gate voltages the data deviate from the model. Particularly notable is the observed departure from the asymptotic high frequency behavior predicted by the model $G(\omega) = B(\omega) \sim \omega^{1/2}$; the experiment shows admittance increasing faster than $\omega^{1/2}$.

We consider several simplifying assumptions of the model of Eqn. (1) that may explain the discrepancies between the model and data. A finite contact resistance $R_C$ is present at the gold electrode/pentacene boundary. However, the effect of $R_C$ in series with the transmission line is to set $1/R_C$ as the upper bound beyond which $G(\omega)$ cannot increase. This is distinct from the observed behavior, in which $G(\omega)$ increases faster than expected rather than being limited at high frequency. The contact impedance itself could be frequency-dependent; in particular, Hamadani, et al.[11] considered a contact capacitance $C_C$, in parallel with $R_C$. However, the combined effect of $R_C$ and $C_C$ always depresses $G(\omega)$ and $B(\omega)$ below the model of Eqn. 1 in a broad range of frequencies centered around $\omega = 1/R_C C_C$, which is in contrast to the experimental observation of enhanced $G(\omega)$ and $B(\omega)$ at high frequencies. Experimentally, we have also performed length-dependent dc measurements on our TFTs, and we find that the resistance varies proportional to the gate length; the contact resistance is smaller than the resistance of a 1 μm-length section of the pentacene film, which agrees with the lack of contact-resistance effects in the ac admittance.

We also consider that the gate capacitance $c$ might not be constant. Interface-trap capacitance (resulting from the movement of the Fermi level through the disordered distribution of electronic states in the organic semiconductor[11]) would reduce $c$. However, Eqn. (1) predicts that for *any* real $\sigma$ and $c$, $G(\omega) = B(\omega)$ in the high-frequency limit. Figure 3 shows the ratio $G(\omega)/B(\omega)$ for the experimental data and for the model of Eqn. (1). The systematic deviation of $G(\omega)/B(\omega)$ to values less than unity at high frequency is a central feature of the data which is unexplained by the model. (Note that including $R_C$ and $C_C$ causes $G(\omega)/B(\omega) > 1$ for $\omega < 1/R_C C_C$ and $G(\omega)/B(\omega) < 1$ for $\omega > 1/R_C C_C$, which is contrary to the experimental data.)

Assuming that the film is spatially uniform, Eqn. (1) is completely general if we assume a complex, frequency-dependent $\sigma$ and $c$. The experimental data, especially the ratio $G(\omega)/B(\omega)$ (Fig. 3), thus indicate a fundamental failure of the assumption that $\sigma$ and $c$ are purely real. However, since the capacitance is dominated by the oxide capacitance (the interface trap capacitance is at most a small correction), it is reasonable to keep the assumption of a purely real and frequency-independent $c$. Then we can relax our assumptions about the transmission line model of Eqn. (1) to allow a complex, frequency-dependent sheet conductivity $\sigma(\omega) = \sigma'(\omega) + j\sigma''(\omega)$, and we can numerically invert Eqn. (2) to obtain $\sigma'(\omega)$ and $\sigma''(\omega)$ directly from a measurement of $G(\omega)$ and $B(\omega)$.

Figure 4 shows $\sigma'(\omega)$ and $\sigma''(\omega)$ for the pentacene TFT calculated from Eqn. (1) allowing $\sigma$ to be complex and frequency-dependent. At low frequency and large negative gate voltage, $\sigma'(\omega) \gg \sigma''(\omega)$, and $\sigma'(\omega)$ is nearly frequency-independent and close to the dc value. However, at higher frequency, or at more positive $V_G$, $\sigma'(\omega)$ rises roughly as a power law. Roughly power-law behavior is also observed in $\sigma''(\omega)$. Once we have calculated $\sigma(\omega)$ we can calculate the characteristic decay length $\lambda = L/z = (8\sigma(\omega)/\omega c)^{1/2}$. We find $\lambda > 100$ μm at all frequencies and gate voltages measured here, which justifies ignoring contact resistance (the contact resistance is less than the resistance of a 1 μm device length). The long decay length also justifies the assumption that the film is uniform; non-uniformities such as grain boundaries have a length scale <1 μm in polycrystalline pentacene.[13]

We now address the question of whether the measured $\sigma'(\omega)$ and $\sigma''(\omega)$ are reasonable for a pentacene thin film. In general, hopping conduction in a disordered medium results in a frequency-dependent conductivity: as the frequency is increased, hopping events occur between localized sites which do not contribute to the dc conductivity but enhance the ac conductivity, which becomes complex.[10] A wide range of disordered materials exhibit a "universal dielectric response" (UDR)[14] in which (roughly) $\sigma'(\omega) = \sigma'(0) + A\omega^s$ and $\sigma''(\omega) = B\omega^s$, with the exponent $s$ dependent on temperature, and typically on order 0.8. The experimental data in Fig. 4 are consistent with UDR over a large portion of the gate voltage and frequency ranges probed, with $0.5 < s < 1$. Further work will be needed to understand the detailed frequency response of the pentacene films.

The frequency dependence of $\sigma(\omega)$ should have profound implications for pentacene devices. In particular, the on/off ratio is severely reduced at high frequency; in our devices, the on/off ratio is reduced from ~$10^5$ at 50 Hz to ~$10^3$ at 20 kHz. Pentacene TFTs will be very lossy in the off state at high frequency. Additionally, the significant susceptivity $\sigma"(\omega)$ will produce an effective parasitic capacitance between source and drain which should be important in ac device models.

In conclusion, we have used a transmission-line technique to measure the frequency-dependent complex conductivity $\sigma(\omega) = \sigma'(\omega) + j\sigma"(\omega)$ of the organic semiconductor pentacene in a thin-film transistor configuration. We find that $\sigma(\omega)$ is strongly frequency dependent, and follows roughly the expectations of universal dielectric response of disordered solids. The technique should be broadly applicable to measure $\sigma(\omega)$ of any thin-film transistor material, and the observation of a complex, frequency-dependent $\sigma(\omega)$ in pentacene has implications for ac applications of pentacene TFTs. We expect that further studies on the frequency-dependent conductivity of single-crystal devices as well as intentionally disordered devices could provide additional insight into the nature of disorder (structural or electronic) that gives rise to the frequency-dependent behavior.

The authors would like to thank Vinod Sangwan for supplying the devices used in this study. This research was supported by the Laboratory for Physical Sciences. The UMD-MRSEC nanofabrication shared equipment facility was used in this work.

Figure Captions

Figure 1: Low-bias dc conductivity of the pentacene thin-film transistor used in this study.

Figure 2: ac conductance $G(\omega)$ and susceptance $B(\omega)$ of the pentacene TFT (solid symbols) measured in the transmission-line configuration (described in text), and the prediction of Eqn. (1) (solid lines) assuming a real, frequency-independent sheet conductivity. From top to bottom, gate voltages are -50, -40, -30, -20, -16, -12, -10, -8, -6, -4 V. Inset shows a schematic of the transmission line measurement configuration.

Figure 3: Ratio of the ac conductance to susceptance, $G(\omega)/B(\omega)$, for the pentacene TFT (solid symbols) measured in the transmission-line configuration (described in text), and the prediction of Eqn. (1) (solid lines) assuming a real, frequency-independent sheet conductivity. From top to bottom, gate voltages are -50, -40, -30, -20, -16, -12, -10, -8, -6, -4 V. (For most gate voltages the solid lines are indistinguishable from a constant value of 1.)

Figure 4: Real and imaginary parts of the sheet conductivity $\sigma'(\omega)$ and $\sigma''(\omega)$ of the pentacene TFT extracted from $G(\omega)$ and $B(\omega)$ using the model of Eqn. (1) allowing for a complex, frequency-dependent sheet conductivity.

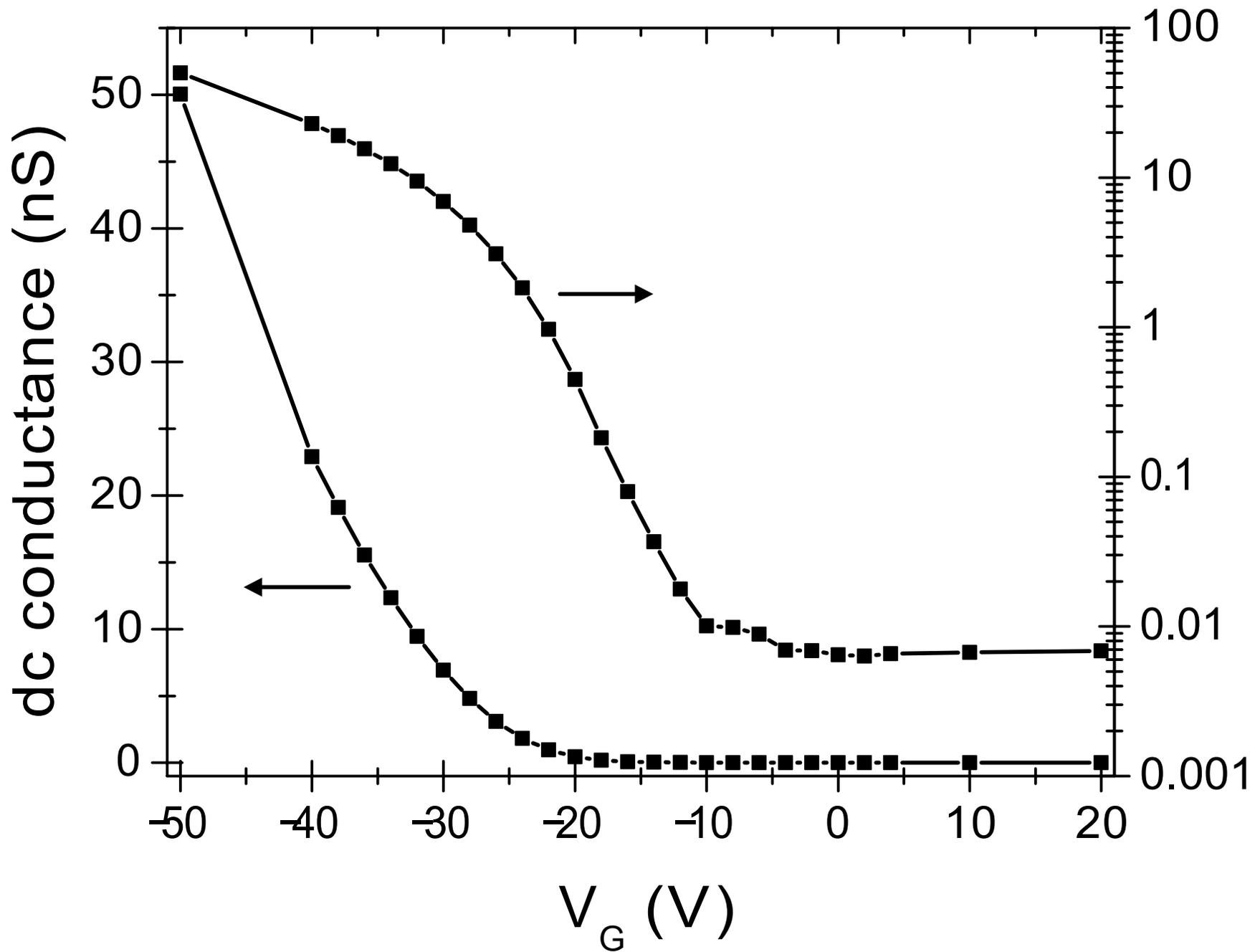

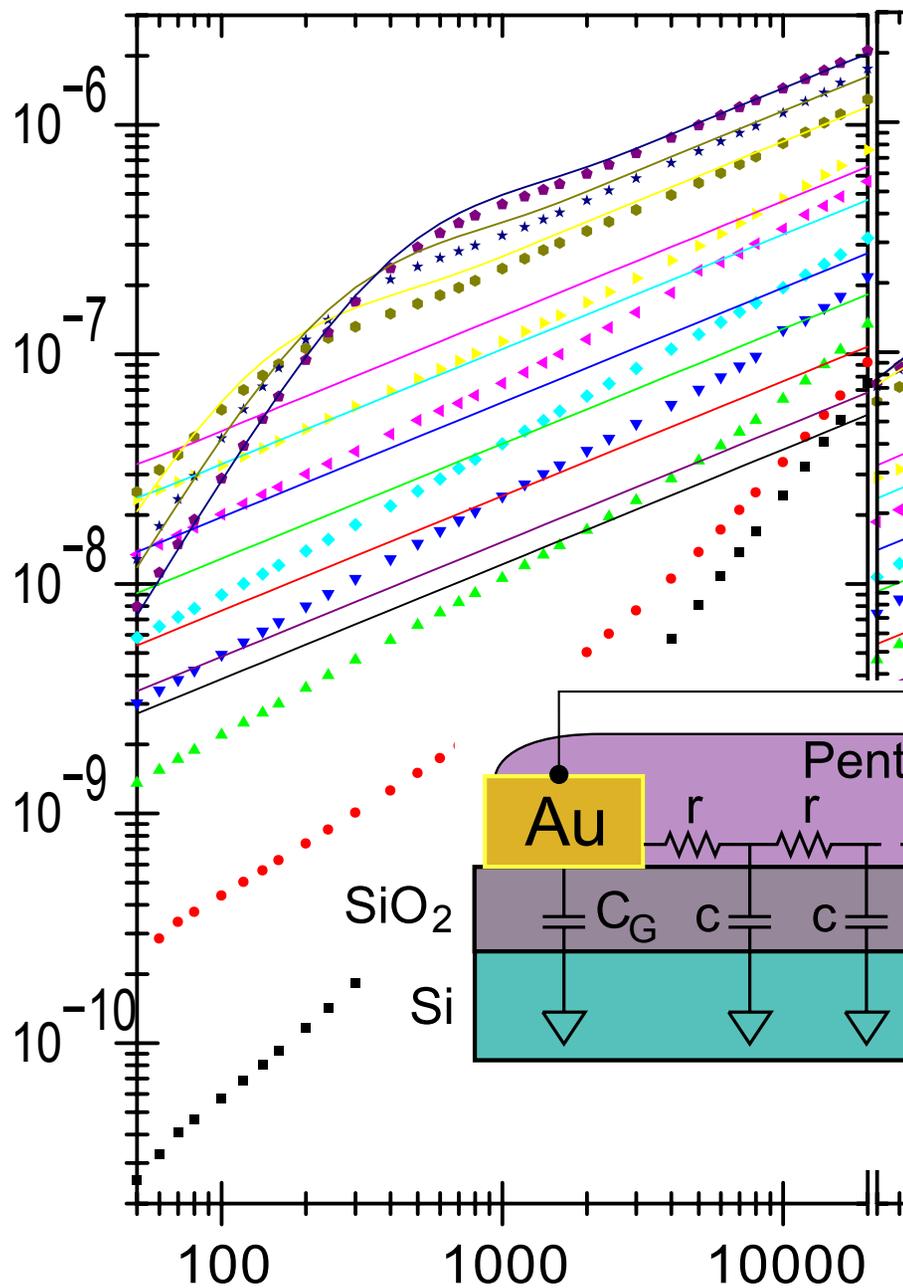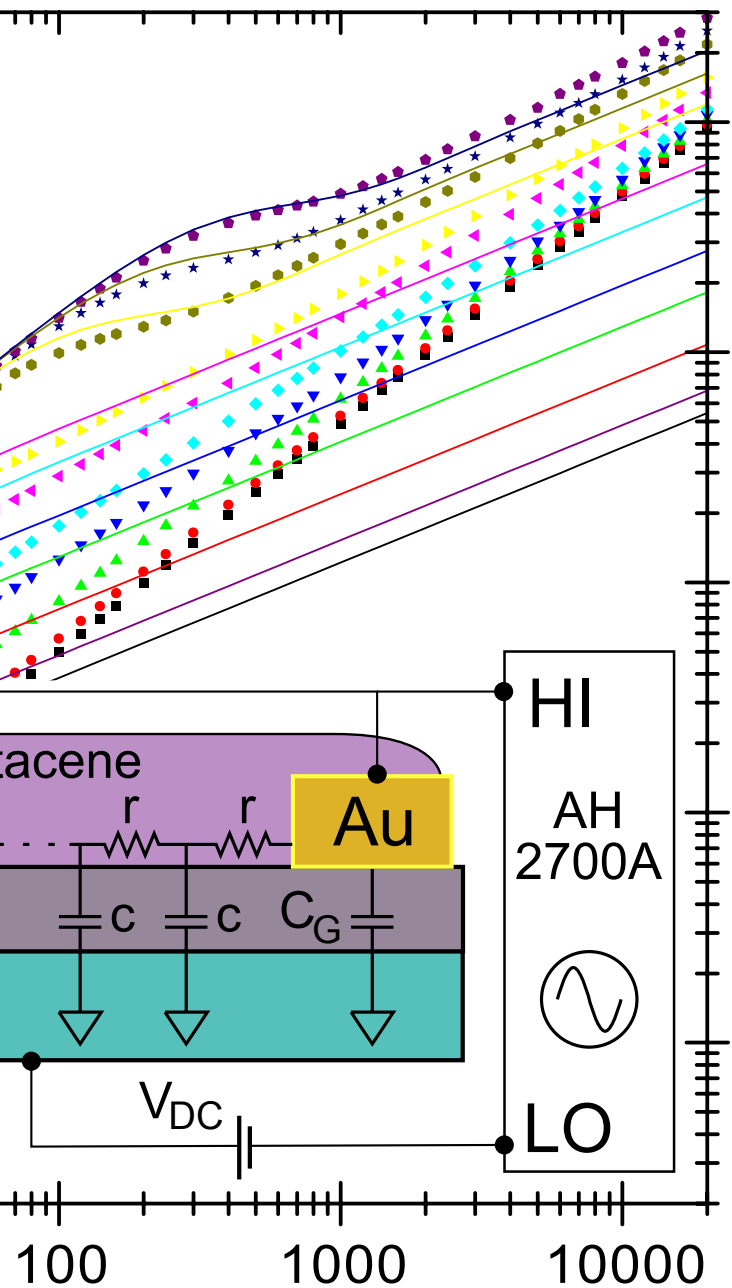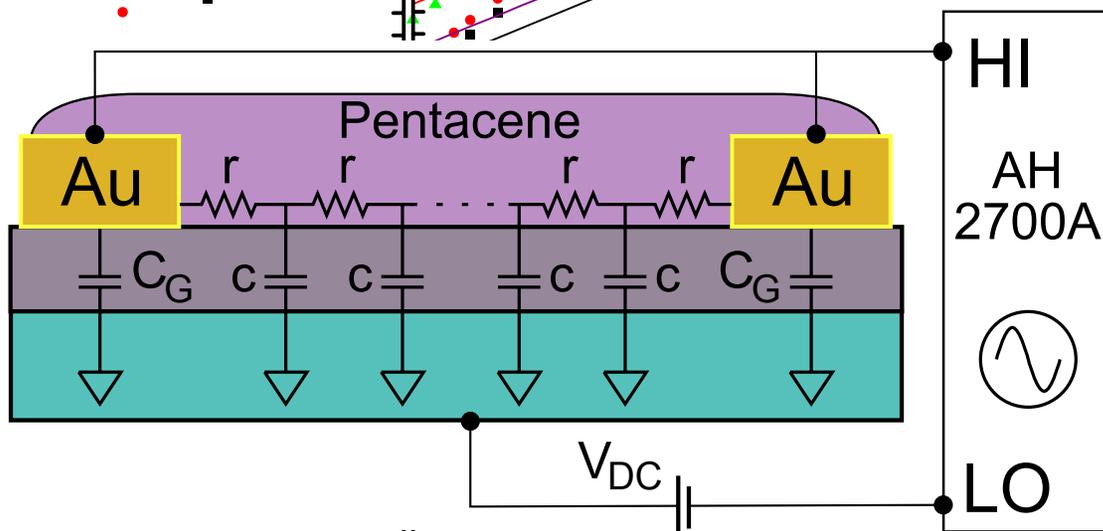

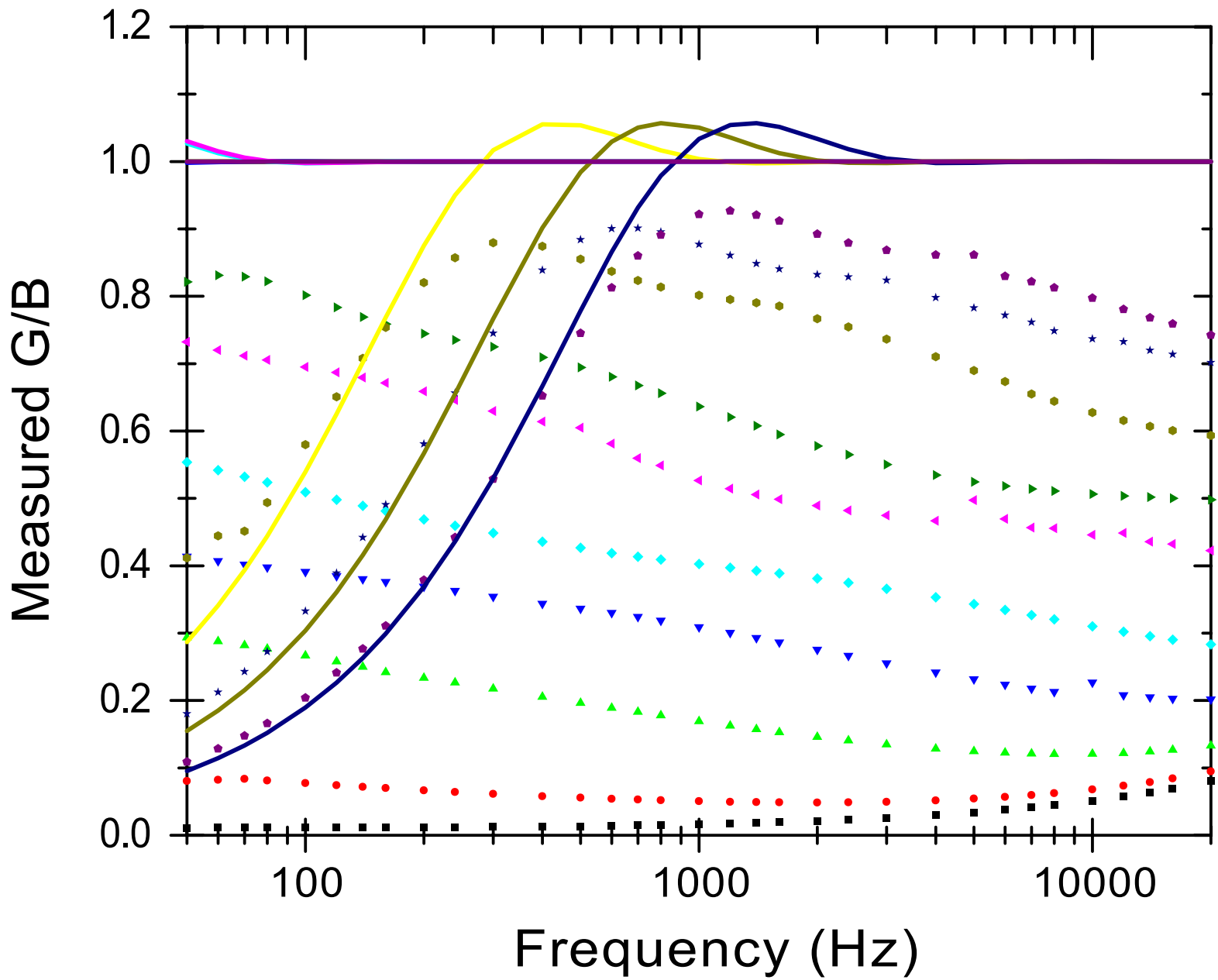

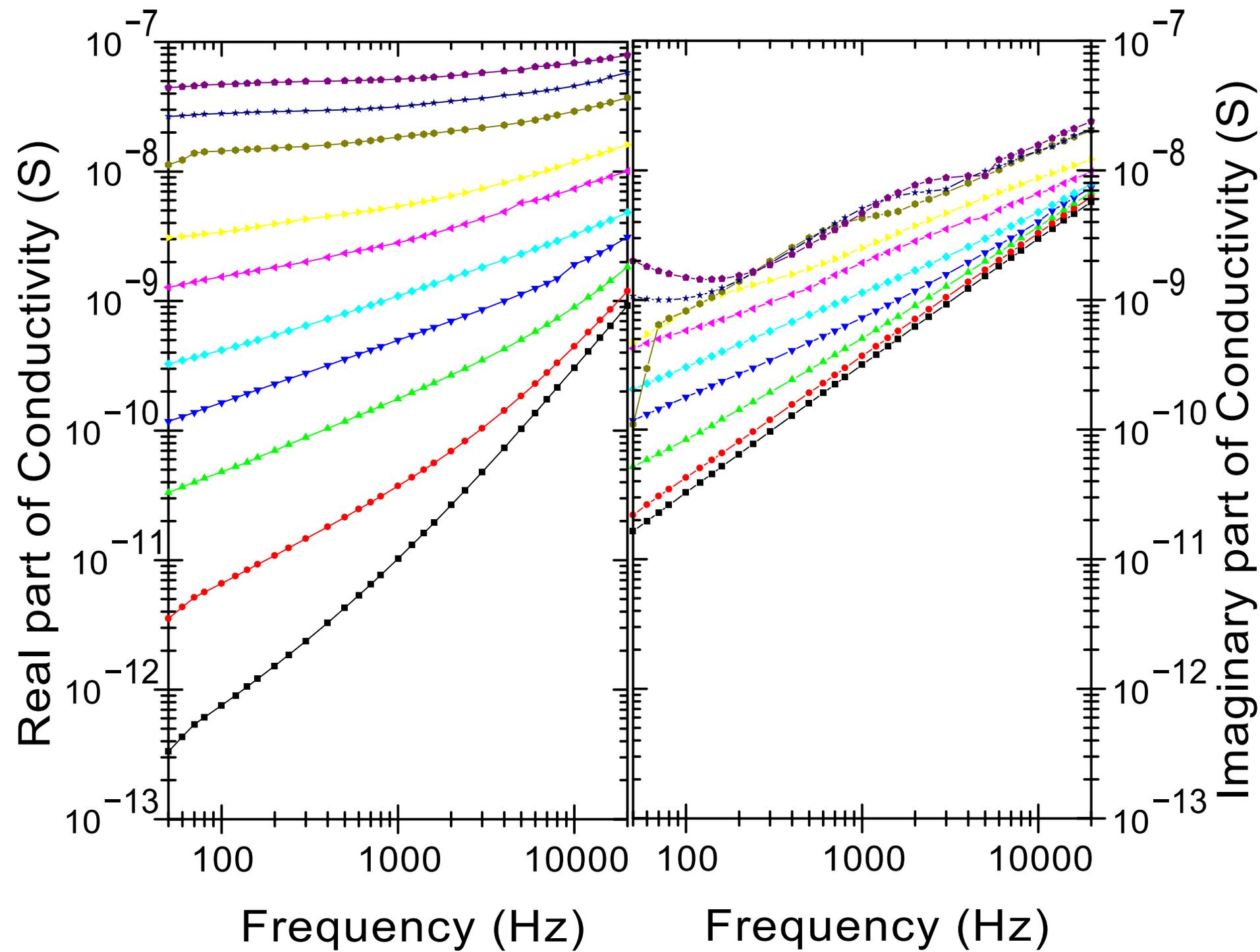